\documentclass[onecolumn,aps,prd,preprintnumbers,showpacs,superscriptaddress,nofootinbib,amsmath,amssymb,floats,floatfix,showkeys,notitlepage,longbibliography]{revtex4-1}

\usepackage{orcidlink}
\usepackage{lipsum}
\usepackage{graphicx}
\usepackage{subfigure}
\usepackage{braket}
\usepackage[utf8]{inputenc}
\usepackage{sans}
\usepackage{changes}
\usepackage{hyperref}
\hypersetup{colorlinks=true,linkcolor=blue,urlcolor=blue,citecolor=blue}
\usepackage{lmodern}

\usepackage{mathrsfs}
\usepackage{mathtools}
\usepackage[toc,page]{appendix}
\usepackage[normalem]{ulem}
\usepackage{adjustbox}
\usepackage{latexsym}
\usepackage{amsmath}
\usepackage{amssymb}
\usepackage{amsfonts}
\usepackage{dcolumn}
\usepackage{bm}
\usepackage{tikz}
\usepackage{bigints}
\usepackage{array,tabularx,multirow,booktabs}
\usepackage[tracking=true]{microtype}
\SetTracking{}{500}
\SetTracking{encoding={*}, shape=sc}{40}
\UseRawInputEncoding %for inputenc error%
\allowdisplaybreaks

\begin{document} 
\title{Exact Taub-NUT-like Black Holes in Einstein-bumblebee gravity: their thermodynamics and thermodynamic topology}

\author{Mustapha Azreg-A\"{\i}nou\orcidlink{0000-0002-3244-7195}}
\email{azreg@baskent.edu.tr}
\affiliation{Ba\c{s}kent University, Engineering Faculty, Ba\u{g}l{\i}ca Campus, 06780-Ankara, T\"{u}rkiye}
\author{Yassine Sekhmani\orcidlink{0000-0001-7448-4579}}
\email{sekhmaniyassine@gmail.com}
\affiliation{Center for Theoretical Physics, Khazar University, 41 Mehseti Street, Baku, AZ1096, Azerbaijan}
\affiliation{Centre for Research Impact \& Outcome, Chitkara University Institute of Engineering and Technology, Chitkara University, Rajpura, 140401, Punjab, India}
\affiliation{Institute of Nuclear Physics, Ibragimova, 1, 050032 Almaty, Kazakhstan}
\affiliation{Department of Mathematical and Physical Sciences, College of Arts and Sciences, University of Nizwa, P.O. Box 33, Nizwa 616, Sultanate of Oman}

\begin{abstract}
We re-derive an exact analytic three-parameter expressions for the non-rotating metric, describing a Taub-NUT-like black hole (BH), and its associated bumblebee field that are solutions to the Einstein-bumblebee gravity. We construct a consistence thermodynamics for the Taub-NUT-like BH and determine its thermodynamic topological class. The Lorentz symmetry breaking affects the mass and temperature of the BH but does not affect its thermodynamic topological classification.
\end{abstract}

\maketitle

\section{Introduction}
Lorentz symmetry constitutes a cornerstone of modern theoretical physics: it underlies relativistic quantum field theory and the Standard Model, and it is integral to the geometric structure of general relativity (GR). Motivated by attempts to reconcile gravity with quantum mechanics and by suggestive features in ultra-high-energy cosmic-ray data, a number of authors have entertained the possibility that Lorentz invariance may be violated at the Planck scale \cite{Colladay:1998fq,Kostelecky:1991ak,Kostelecky:1989jp,Gambini:1998it,Carroll:2001ws}. Crucially, even minute Planck-scale violations can leave low-energy effective imprints on spacetime dynamics; by modelling these remnants and confronting them with high-precision astrophysical and laboratory observations, one obtains a powerful probe of candidate quantum-gravity scenarios. Such an enterprise therefore requires both (i) a controlled derivation of the low-energy signatures of Lorentz-symmetry breaking and (ii) careful, model-aware comparison with observational data to determine whether any putative signal is compatible with a consistent quantum-gravity framework.

The Bumblebee model is a vector-tensor extension of Einstein-Maxwell theory in which a dynamical vector field $B_\mu$ (the bumblebee field) is nonminimally coupled to the gravitational sector. Under an appropriate potential the field can acquire a nonzero vacuum expectation value (VEV), thereby triggering spontaneous breaking of local Lorentz symmetry and endowing the spacetime with a preferred direction. This symmetry breaking generically produces anisotropic stress-energy contributions that modify the standard GR dynamics. Kosteleck\'{y} and Samuel originally proposed the bumblebee framework to investigate spontaneous Lorentz violation in gravity \cite{Kostelecky:1989jw}, and Casana et al. subsequently constructed a Schwarzschild-like bumblebee BH solution \cite{Casana:2017jkc}. A range of additional spherically symmetric configurations has since been reported, including wormholes, global monopoles, solutions featuring an effective cosmological constant, and extensions that incorporate Einstein-Gauss-Bonnet corrections \cite{Maluf:2020kgf,Ovgun:2018xys,Gullu:2020qzu,Ding:2021iwv,Adailton23}. Rotating counterparts have also been explored in~\cite{Poulis:2021nqh,Adailton24}. Motivated by these developments, the imprint of Lorentz-symmetry breaking in strong-gravity environments has become a lively and expanding subfield of BH physics \cite{Liu:2022dcn,Mai:2023ggs,Yang:2023wtu,Xu:2023xqh,Zhang:2023wwk,Wang:2021gtd}.

The Taub-NUT spacetime \cite{Taub:1950ez, Newman:1963yy} is a key exact solution of Einstein's field equations, representing the simplest nontrivial smooth deformation of the Schwarzschild manifold through the introduction of a single additional parameter, viz. the NUT (Newman-Unti-Tamburino) charge $n$. This parameter endows the geometry with a gravitomagnetic monopole character and gives rise to a rich causal and topological structure. The Taub-NUT geometry, unlike many BH spacetimes, does not exhibit a curvature scalar singularity. Instead, it features a line-like singularity along the polar axis, known as the Misner string. This singularity encodes the nontrivial gravitomagnetic structure of the spacetime and has significant implications for the global causal and topological properties of the manifold~\cite{Misner:1963fr}. The thermodynamic behaviour of Taub-NUT black holes (BHs) has garnered significant analytical and numerical attention (see, e.g., \cite{eq1,eq2,eq3,Wu:2019pzr,Liu:2022wku,eq4,Chen:2024knw}). Recent breakthroughs in observational astronomy, particularly in the realms of gravitational-wave detection~\cite{LIGOScientific:2017vwq,LIGOScientific:2017zic} and horizon-scale imaging conducted by the Event Horizon Telescope collaboration~\cite{EventHorizonTelescope:2019dse,EventHorizonTelescope:2019uob,EventHorizonTelescope:2019jan,EventHorizonTelescope:2019ths,EventHorizonTelescope:2019pgp,EventHorizonTelescope:2019ggy}, have considerably broadened the empirical dataset pertaining to compact objects. These advancements have also spurred targeted searches for observable signatures related to NUT charge. The expanding body of literature has investigated whether astrophysical BHs, such as M87* and Sgr A*, might possess a nonzero NUT charge~\cite{Chakraborty:2019rna,Ghasemi-Nodehi:2021ipd,Jafarzade:2025zbg}. Complementary indications have been explored through X-ray binary spectral analyses~\cite{Chakraborty:2017nfu}, as well as in proposals suggesting that primordial BHs, when evaluated as candidates for dark matter, could also have a NUT charge~\cite{Chakraborty:2022ltc,Chakraborty:2023wpz}. These theoretical and observational developments collectively elevate the Taub-NUT solution from being merely a mathematical curiosity to a model that is physically grounded and has concrete, testable implications for astrophysics.

In Sec.~\ref{secebg} we provide a concise introduction to Einstein-bumblebee gravity and re-derive an exact three-parameter non-rotating Taub-NUT-like BH~\cite{Chen25}. In Sec.~\ref{secther} we construct a consistent BH thermodynamic of the solution and provide expressions for its thermodynamic entities, which generalize existing ones. Sec.~\ref{sectop} is devoted to the thermodynamic topological classification. After a short introduction to Poincar\'e-Hopf Index Theorem for manifolds with or without boundary, we analyze the classification of our three-parameter solution based on the version of the Theorem for manifolds without boundary. We conclude in Sec.~\ref{seccon}.

\section{Einstein-bumblebee gravity\label{secebg}}
The action describing Einstein-bumblebee gravity with one vector field $B^{\mu}$, the bumblebee field, is given by~\cite{Casana:2017jkc}
\begin{equation}\label{action}
I=\int d^{4}x\sqrt{-g}[\frac{1}{2\kappa}(R+\gamma B^{\mu}B^{\nu}R_{\mu\nu})-\frac{1}{4}B_{\mu\nu}B^{\mu\nu}-V]\,.
\end{equation}
The bumblebee field is non-minimally coupled to gravity via the Ricci tensor $R_{\mu\nu}$. Here $\gamma$ is a real coupling constant that describes the non-minimal coupling and $B_{\mu\nu}$ is the bumblebee field strength defined by $B_{\mu\nu}=\partial_{\mu}B_{\nu}-\partial_{\nu}B_{\mu}$. Lorentz symmetry is spontaneously broken only if $B_{\mu}$ has a nonzero vacuum expectation value (VEV): $<B_{\mu}>=b_{\mu}$. The potential $V$ is function of $(B_{\mu}B^{\mu}\pm b^2)$ and it vanishes if $B^\mu =b^\mu$. If the derivative of $V$ with respect to its argument also vanishes at $B^\mu =b^\mu$, this implies $b^{\mu}b_{\mu} = \mp b^2$, leading to conclude that the VEV $b^\mu$ is either timelike or spacelike.

From now on, we assume that the bumblebee field is frozen in its VEV $b^{\mu}$. This means that both $V$ and its derivative vanish. In this case, the field equations take the simple form
\begin{equation}\label{EOM-gr}
F_{\mu\nu}\equiv R_{\mu\nu}-\frac{1}{2}g_{\mu\nu}R-\kappa T_{\mu\nu}^{\text{(Bee)}}=0,
\end{equation}
with
\begin{equation}\label{SETB}
\begin{split}
	T_{\mu\nu}^{\text{(Bee)}}=&\;b_{\mu\alpha}b^{\alpha}_{~\nu}-\frac{1}{4}b_{\alpha\beta}b^{\alpha\beta}+\frac{\gamma}{\kappa}\Big[\frac{1}{2}b^{\alpha}b^{\beta}R_{\alpha\beta}g_{\mu\nu}-b_{\mu}b^{\alpha}R_{\alpha\nu}-b_{\nu}b^{\alpha}R_{\alpha\mu}\\&+\frac{1}{2}\nabla_{\alpha}\nabla_{\mu}(b^{\alpha} b_{\nu})
	+\frac{1}{2}\nabla_{\alpha}\nabla_{\nu}(b^{\alpha} b_{\mu})-\frac{1}{2}\nabla^2(b_{\mu}b_{\nu})-\frac{1}{2}g_{\mu\nu}\nabla_{\alpha}\nabla_{\beta}(b^{\alpha}b^{\beta})\Big]\,
\end{split}
\end{equation}
and 
\begin{equation}\label{EOM-bee}
E_{\nu}\equiv \nabla^{\mu}b_{\mu\nu}+\frac{\gamma}{\kappa}b^{\mu}R_{\mu\nu}=0.
\end{equation}

For constructing a non-rotating Taub-NUT type solution in  Einstein-Bumblebee theory, we follow~\cite{Chen25} and we assume a metric of the Taub-NUT form with two arbitrary functions $h(r),f(r)$
\begin{equation}\label{ansatz}
ds^2=-h(r)(dt+2n\cos\theta d\phi)^2+\frac{dr^2}{f(r)}+(r^2+n^2)(d\theta^2+\sin^2\theta d\varphi^2)\,.
\end{equation}
We take the VEV $b^{\mu}$ with one radial component 
\begin{equation}\label{ansatzb}
b_{\mu}=(0,b(r),0,0).
\end{equation}
Since $b^{\mu}b_{\mu}=\mp b_0^2=\text{const}$, the bumblebee field $b(r)$ can be directly derived
\begin{equation}
b^{\mu}b_{\mu}=g^{rr}b(r)^2=\mp b_0^2~\rightarrow ~b(r)=\sqrt{\frac{\mp b_0^2}{f(r)}}.
\end{equation}
Ansatz~\eqref{ansatzb} implies $b_{\mu\nu}=0$ and, consequently, the first two terms in~\eqref{SETB} vanish identically.

Another asnatz, which constituted the essence of the work done in~\cite{Chen25}, is to look for solution where the functions $f$ and $h$ are proportional, and this is inspired from the case of the Schwarzschild-like black hole solution derived in~\cite{Casana:2017jkc}
\begin{equation}\label{ansatz2}
f(r)=K\, h(r).
\end{equation}

To solve the field Eqs.~\eqref{EOM-gr} and~\eqref{EOM-bee}, we adopt the ($t,\,r,\,u,\,\varphi$) coordinates instead of the usual ones ($t,\,r,\,\theta,\,\varphi$), where $u=\cos\theta$. In these coordinates the metric~\eqref{ansatz} takes the form
\begin{equation}\label{ansatz3}
ds^2=-h(r)(dt+2nu d\phi)^2+\frac{dr^2}{Kh(r)}+(r^2+n^2)\Big[\frac{du^2}{1-u^2}+(1-u^2) d\varphi^2\Big].
\end{equation}
With the ansatzes~\eqref{ansatz2} and~\eqref{ansatz3}, all the components of the l.h.s of the bumblebee field equation~\eqref{EOM-bee}, but $E_r$, vanish identically:
\begin{equation}\label{f1}
E_\nu=\bigg(0,\,\frac{K b_0 \gamma }{2 \kappa  (n^2+r^2)^2 \sqrt{f(r)}}\, \Big[4 n^2 h+(n^2+r^2) [2 r h'+(n^2+r^2)h'']\Big],\,0,\,0\bigg)	\,.
\end{equation}
Equations~\eqref{EOM-bee} will be satisfied, that is, $E_\nu =(0,\,0,\,0,\,0)$, if we take
\begin{equation}\label{h}
h(r)=\frac{r^2-2mr-n^2}{r^2+n^2}\,,
\end{equation}
which is a solution to $E_r =0$, that is, to the differential equation $4 n^2 h+(n^2+r^2) [2 r h'+(n^2+r^2)h'']=0$. With this expression of $h(r)$, the non-vanishing components of $F_{\mu\nu}$~\eqref{EOM-gr} are
\begin{align}
&F_{t t}=-\frac{2 [K (1+\ell )-1] n^2 [n^2+(2 m-r) r]^2}{(n^2+r^2)^4}\,,\\
&F_{t \varphi }=-\frac{4 [K (1+\ell )-1] n^3 u [n^2+(2 m-r) r]^2}{(n^2+r^2)^4}\,,\\
&F_{u u}=\frac{[K (1+\ell )-1] [n^4+4 n^2 (m-r) r-r^4]}{(1-u^2) (n^2+r^2)^2}\,,\\
&F_{\varphi  \varphi }=\frac{[K (1+\ell )-1] \{(n^2+r^2)^2 [n^4+4 n^2 (m-r) r-r^4]- u^2 [3 n^4+2 n^2 (4 m-3 r) r-r^4]
	(3 n^4+4 m n^2 r+r^4)\}}{(n^2+r^2)^4}\,,
\end{align}
they all vanish if
\begin{equation}
K=\frac{1}{1+\ell}\,,
\end{equation}
where $\ell =\gamma b_0^2$. We have thus obtained the exact Taub-NUT-like solution
\begin{equation}\label{sol}
ds^2=-\frac{r^2-2mr-n^2}{r^2+n^2}(dt+2n\cos\theta d\varphi)^2+\frac{(1+\ell)(r^2+n^2)}{r^2-2mr-n^2}~dr^2+(r^2+n^2)(d\theta^2+\sin^2\theta d\varphi^2)\,,
\end{equation}
which reduces to the Schwarzschild-like BH derived in~\cite{Casana:2017jkc} if $n=0$, to the Taub-NUT solution if $\ell =0$, and to Schwarzschild BH if both $n=0$ and $\ell =0$. This metric was first derived in~\cite{Chen25}.

For arguments emphasized in~\cite{Chen25}, it is not possible to set $h(r)=f(r)$ by rescaling the time coordinate $t\rightarrow t'=\sqrt{1+\ell}\,t$.

\section{Thermodynamics\label{secther}}
There are many equivalent formulas for the determination of the thermodynamic mass $M$ in terms of the mass parameter $m$~\cite{mass1,mass2,mass3}. If no anti-de Sitter term, we use the formula given in~\cite{mass3}, which applies to $d$-dimensional spacetimes
\begin{equation}\label{ms1}
M = \frac{(d-2)V_{d-2}}{16\pi (d-3)}\lim_{r\to\infty}\Big(|g_{tt}'(r)||g_{\theta\theta}|^{(d-2)/2}~\frac{1}{\sqrt{|g_{tt}g_{rr}|}}\Big),\qquad V_{d-2}=\frac{2~\pi ^{(d-1)/2}}{\Gamma\big(\frac{d-1}{2}\big)}.
\end{equation}
Here $V_{d-2}$ is the volume of the $(d-2)$-dimensional unit sphere. Taking $d=4$, $V_2=4\pi$ [$\Gamma(3/2)=\sqrt{\pi}/2$], $g_{tt}(r)=-h(r)$~\eqref{h}, $g_{rr}=(1+\ell)/h(r)$ (note that $|g_{tt}g_{rr}|=1/K=1+\ell$), and $g_{\theta\theta}=r^2+n^2$ we obtain
\begin{equation}\label{M}
M = \frac{m}{\sqrt{1+\ell}}\,.
\end{equation}
The temperature of the BH~\eqref{ansatz} is expressed as
\begin{equation}\label{T}
T = \frac{1}{4\pi}\sqrt{\frac{|g_{tt}|}{|g_{rr}|}}~\frac{|g_{tt}|'}{|g_{tt}|}\bigg|_{r=r_+} = \frac{1}{4\pi}\sqrt{\frac{h(r)^2}{1+\ell}}~\frac{h(r)'}{h(r)}\bigg|_{r=r_+}=\frac{h(r)'|_{r=r_+}}{4\pi\sqrt{1+\ell}}=\frac{1}{4\pi\sqrt{1+\ell}~r_+}\,,
\end{equation}
where the r.h.s is evaluated at the unique horizon $r_+\;(=r_h)$ defined by $h(r_+)=f(r_+)=0$,
\begin{equation}\label{oh}
r_+^2 - 2mr_+ - n^2 = 0\to r_+ =m+\sqrt{m^2+n^2}\to m=\frac{r_+^2 - n^2}{2r_+}\,.
\end{equation}

The thermodynamics of Taub-NUT-(dS or AdS) has some equivalent descriptions~\cite{eq1,eq2,eq3,Wu:2019pzr,Liu:2022wku,eq4,Chen:2024knw}. Since we are not considering dS and adS terms in our investigation, we find the description given in~\cite{eq3} the more convenient one and we will adopt it along with the expression of the entropy
\begin{equation}\label{entropy}
S=\pi (r_+^2 + n^2)\,.
\end{equation}

The fact that some parameters may play the role of extra charges or order parameters is not new in BH thermodynamics~\cite{Poincare,Jiayue23}, provided the associated thermodynamic state variable is not a redundant variable~\cite{eq2}. In this spirit, in Ref.~\cite{eq3}, a charge $N=-4\pi n^3/r_+$ and a potential $\psi=1/(8\pi n)$ have been associated to the NUT parameter $n$ in such a way that $M=2(TS+\psi N)$, where $T=1/(4\pi r_+)$~\cite{eq3}. Since in our case, the temperature has been multiplied by a factor of $1/\sqrt{1+\ell}$, it is the potential $\psi$ that has received the same factor. Hence, the generalized ($\psi,\,N$) take the form
\begin{equation}\label{pN}
\psi =\frac{1}{8\pi n\sqrt{1+\ell}}\,,\qquad N=-\frac{4\pi n^3}{r_+}\,.	
\end{equation}
[Note that $N$ depends on ($S,\,n$)~\eqref{entropy}, and so it is not a redundant variable]. Note that since the Taub-NUT-like metric~\eqref{sol} is continuous for all allowed values of $\ell$ [all known metrics are obtainable from metric (17) for specific values of $\ell$], so is the thermodynamics of this class of BHs. By continuity of the parameter $\ell$, we have just extended the thermodynamic analysis made in~\cite{eq3} to the Taub-NUT-like BHS.

Using~\eqref{M}, \eqref{T}, \eqref{entropy} and~\eqref{pN} it is straightforward to obtain
\begin{equation}\label{thermo1}
M=2(TS+\psi N)\to \text{d}M = T \text{d}S + \psi \text{d}N\,.
\end{equation}

Using~\eqref{oh}, the expression of $M$~\eqref{M} also reaches the form
\begin{equation}\label{thermo2}
M(S,N) = \frac{\sqrt{S-\pi n^2}}{2\sqrt{\pi}\sqrt{1+\ell}} + \frac{N}{8\pi n\sqrt{1+\ell}}\,,
\end{equation}
where $S$ and $N$ are taken as independent variables. This immediately yields $\text{d}M = T \text{d}S + \psi \text{d}N$ since $\partial_S M\big|_N=T$~\eqref{T} and  $\partial_N M\big|_S=\psi$~\eqref{pN}.

\section{Thermodynamic topological classes\label{sectop}}
In thermodynamic topology~\cite{ettbh}, we consider a two-variable function  $\tilde{\mathcal{F}}(r_h,\Theta)$ that is sum of the off-shell free energy $F(r_h)$, which depends on the horizon location $r_+ = r_h$, and some function $G(\Theta)$ of $\Theta$
\begin{equation}\label{F0}
\tilde{\mathcal{F}}(r_h,\Theta)=F(r_h)	+ G(\Theta)\,.
\end{equation}
In the black hole literature, there are two ways to choose the function $G(\Theta)$ as this is related to the application of the Poincar\'e-Hopf Index Theorem for manifolds with or without boundary~\cite{Poincare1,Poincare2,Hopf,book}. In this work, as was the case in~\cite{AR25}, we apply the theorem for manifolds without boundary, which states: \\

\noindent \textsf{If $\phi$ is a vector field with only isolated critical points (CPs) on a surface $\mathcal{M}$ that can be orientable or nonorientable, and if $X_1$, $X_2$, ..., $X_i$ is the set of all isolated CPs (zeros) of $\phi$, then $\sum _{i}\text{ind}(\phi,X_i)=\chi(\mathcal{M})$, where $\chi(\mathcal{M})$ is the Euler characteristic of $\mathcal{M}$}.\\

\noindent Here $\text{ind}(\phi,X)\in \mathbb{Z}$ is the index (or winding number) of $\phi$ at the isolated zero or CP $X$, $\phi (X) = 0$, which is the change in angle that the unit vector $\phi /||\phi||$ makes after traveling counterclockwise once around any small closed path that surrounds the CP $X$~\cite{AR25}.

Within the requirements of the above-mentioned theorem for manifolds without boundary, there are many equivalent choices for the function $G(\Theta)$. Following~\cite{AR25}, we make the choice
\begin{equation}\label{F1}
G(\Theta)=\frac{\sin 2\Theta}{2}\,,\qquad 0\leq\Theta\leq\pi \,.
\end{equation}
With this choice, the surface $\mathcal{M}$, defined by $r_m\leq r_h<\infty$ ($r_m$ is the smallest value of $r_h$) and $0\leq\Theta\leq\pi$ has the topology of a cylinder with no boundary (see Fig.~1 of~\cite{AR25}). Note that the CPs of $\tilde{\mathcal{F}}(r_h,\Theta)$ or $G(\Theta)$ all lie on the lines $\Theta=3\pi/4$ and $\Theta=\pi/4$. For topological classification, we define the topological charge $Q$ of the black hole (or the total winding number $W$, as denoted in some literature) as the sum of the indices of all CPs on the line $\Theta=3\pi/4$:
\begin{equation}\label{F2}
Q=W=\sum _{i}^{(\Theta=3\pi/4)}\text{ind}(\phi,X_i)\,.
\end{equation}
Since the Euler characteristic of a cylinder $\chi(\mathcal{M})=0$, the sum of the indices of all CPs on the line $\Theta=\pi/4$ is $-Q$, so that the total sum of all indices is $Q+(-Q)=\chi(\mathcal{M})=0$.

The thermodynamic topological classification of the Taub-NUT-like BH in Einstein-bumblebee gravity is not difference from that of the Lorentzian Taub-NUT Black Hole~\cite{AR25}. The black hole mass $M$ and other thermodynamical quantities required for topological classification are regrouped here
\begin{equation}~\label{here}
M=\frac{m}{\sqrt{1+\ell}}=\frac{r_h^2 - n^2}{2r_h\sqrt{1+\ell}}\,, \quad S=\pi \left(r_h^2+n^2\right), \quad  N=-\frac{4\pi n^3}{r_h}, \quad \psi= \frac{1}{8\pi n\sqrt{1+\ell}},\quad T= \frac{1}{4\pi\sqrt{1+\ell}~r_h}\,,
\end{equation}
where Eq.~\eqref{oh} has been used. Here $S$ and $T$ represent the entropy and Hawking temperature, whereas $N$ represents the gravitational Misner charge and $\psi$ the corresponding associated potential. We see that the temperature $T$ has the following limits
\begin{equation}\label{TT}
T(r_m)=\infty \quad \text{and} \quad T(\infty)=0,
\end{equation}
where $r_m=0$ for the Taub-NUT-like BH in Einstein-bumblebee gravity. Conditions~\eqref{TT} lead to conclude that this BH has the thermodynamic properties of Class I~\cite{AR25,class}.
\begin{equation}\label{class1}
\text{Class I: }Q^{1-}=[\uparrow,\uparrow]=[-,-]\,.
\end{equation}

\begin{figure}[!htb]
\centering
\includegraphics[width=0.7\textwidth]{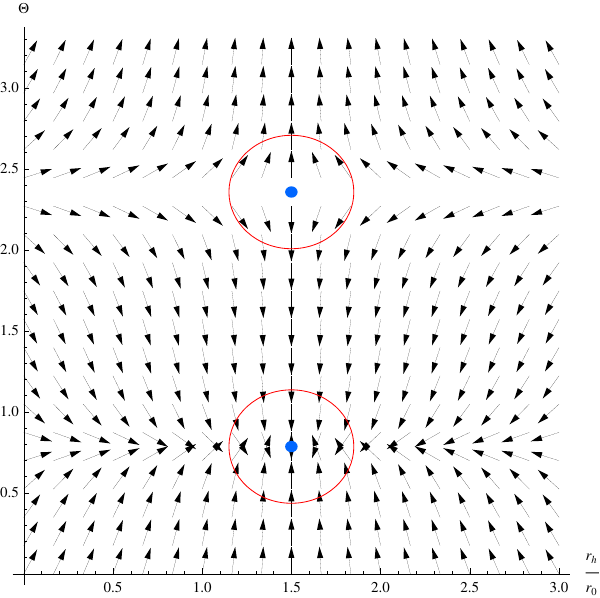}
\caption{Vector field plots of the unit vector field $\phi/||\phi||$ for the Taub-NUT-like solution~\eqref{sol} taking $\tau=6\pi r_0\sqrt{1+\ell}$, where $r_0$ represents an arbitrary length scale defined by the size of a cavity surrounding the black hole. The blue dots represent CPs located at $(r/r_0,\Theta)=(3/2,\pi/4),\,(3/2,3\pi/4)$. For the upper CP (corresponding to $\Theta =3\pi /4$), as one moves counterclockwise once around the red closed curve, the vector field rotates clockwise $2\pi$ radians, so that the index or the winding number is $-1$.}
\label{FigTNUT}
\end{figure}
\begin{figure}[!htb]
\centering
\includegraphics[width=0.5\textwidth]{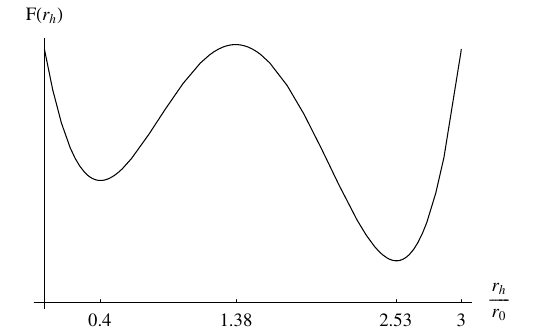}
\caption{$F(r_h)$ versus $r_h$ where $F(r_h)$ has three extreme values. The figure depicts the case where the two arrows, as those shown in~\eqref{class1}, are both downward. As one moves on the line $\Theta =3\pi /4$ from $r=r_m$ to $\infty$, the first extreme value of $F(r_h)$ that one meets is a minimum (occurring at $r_h=0.4 r_0$), so the first arrow is downward. Now, as one moves on the line $\Theta =3\pi /4$ from  $\infty$ to $r=r_m$, the first extreme value of $F(r_h)$ that one meets is again a minimum (occurring at $r_h=2.53 r_0$), so the second arrow is also downward. We do not care about the nature of intermediate extreme value, as the maximum value occurring at $r_h=1.38 r_0$.}
\label{Figarrows}
\end{figure}

To show how the conclusion~\eqref{class1} has been drawn based on the method developed in~\cite{AR25}, consider the on-shell free energy 
\begin{equation}
F=M-ST-\psi N\,.
\end{equation}
The corresponding off-shell free energy takes the form
\begin{equation}\label{off}
\tilde{\mathcal{F}}=F(r_h)+G(\Theta)=\frac{r_h}{2\sqrt{1+\ell}}-\frac{\pi (r_h^2+n^2)}{\tau}+\frac{\sin 2\Theta}{2},
\end{equation}
where from~\eqref{here} we have $M-\psi N =r_h/(2\sqrt{1+\ell})$. Here $1/\tau$ is the temperature of the cavity that surrounds the black hole~\cite{York}. The $\phi$-vector field is the gradient of $\tilde{\mathcal{F}}$
\begin{eqnarray}\label{LTNUT2}
\phi^{r_h}=\frac{1}{2\sqrt{1+\ell}}-\frac{2\pi r_h }{\tau}, \quad \phi^{\Theta}=\cos 2\Theta\,.
\end{eqnarray}
We look for the CPs of $\tilde{\mathcal{F}}$, which correspond to $\phi^{r_h}=0$ and $\phi^{\Theta}=0$. The latter equation yields two solutions in the interval $0\leq\Theta\leq\pi$ and we choose the solution $\Theta =3\pi /4$. The other equation, $\phi^{r_h}=0$, yields
\begin{equation}\label{CP}
r_h=\frac{\tau}{4\pi\sqrt{1+\ell}}\,.
\end{equation}
It is straightforward to show that $r_h=\tau/(4\pi\sqrt{1+\ell})$ yields a maximum of $F(r_h)$ [and the CP ($\tau/(4\pi\sqrt{1+\ell}),\,3\pi /4$) yields a maximum of $\tilde{\mathcal{F}}$]. The left arrow in~\eqref{class1} is to mean that as one moves on the line $\Theta =3\pi /4$ from $r=r_m$ to $\infty$, the first extreme value of $F(r_h)$ that one meets is a maximum; the right arrow in~\eqref{class1} is to mean that as one moves on the line $\Theta =3\pi /4$ from  $\infty$ to $r=r_m$, the first extreme value of $F(r_h)$ that one meets is a maximum. This is true for $F(r_h)$ as it has only one maximum, so moving leftward or rightward one meets one and only one extreme value, which is a maximum. In case where $F(r_h)$ has many extreme values, as in Fig.~\ref{Figarrows}, the direction of the arrows is determined by the nature of the first extreme value that one meets as one moves leftward (from $r=r_m$ to $\infty$) and rightward (from  $\infty$ to $r=r_m$): If the first extreme value from the left is a maximum, we insert an upward arrow in the first position, and if it is a minimum (as in Fig.~\ref{Figarrows}), we insert a downward arrow in the first position. In the second position, we insert an upward or a downward arrow if the first extreme value from the right is a maximum or a minimum (as in Fig.~\ref{Figarrows}). In this approach, we do not care about the nature of intermediate extreme values, as the local maximum in Fig.~\ref{Figarrows}. In~\cite{class}, plus and minus signs have been used instead of arrows [as $[-,-]$~\eqref{class1}], where an upward arrow corresponds to minus sign and a downward arrow corresponds to plus sign.

In~\cite{AR25} we established the rules:
\begin{align}
&\text{For }G(\Theta)=\frac{\sin 2\Theta}{2}\,,\nonumber\\
&\text{If }F(r_h)=\text{ max},\quad X_1=(r_h,\,\pi/4),\quad  X_2=(r_h,\,3\pi/4)\Longrightarrow \nonumber\\
\label{r1}&\text{ind}(\phi,X_1)=1 \text{ and }\text{ind}(\phi,X_2)=-1\,,\\
&\text{If }F(r_h)=\text{ min},\quad X_1=(r_h,\,\pi/4),\quad  X_2=(r_h,\,3\pi/4)\Longrightarrow \nonumber\\
\label{r2}&\text{ind}(\phi,X_1)=-1 \text{ and }\text{ind}(\phi,X_2)=1\,.
\end{align}
The value of $-1$ in~\eqref{class1} is the total winding number, the evaluation of which is based on the rule~\eqref{r1}: since $ X_2=(r_h,\,3\pi/4)$ corresponds to a maximum of $F(r_h)$, the associated winding number is $-1$ and, since there is only one CP as one moves from $r_m$ to $\infty$ (along the line $\Theta =3\pi /4$), the total winding number is $-1$. In our approach supporting plots are not necessary, however, we attached Fig.~\ref{FigTNUT} for clarifications: For the upper CP (corresponding to $\Theta =3\pi /4$), the change in the angle that the unit vector $\phi /||\phi||$ makes after traveling counterclockwise once around the red closed curve is $-2\pi$ so that the index or the winding number is $-1$.

We conclude that the Lorentz symmetry breaking does not affect the thermodynamic topological classification of the Taub-NUT-like BHs in Einstein-bumblebee gravity, that is, the thermodynamic topological classification of the Taub-NUT-like BH in Einstein-bumblebee gravity is not difference from that of the Lorentzian Taub-NUT BH~\cite{AR25}.

\section{Conclusion\label{seccon}}
We have added a three-parameter family of exact solutions to the open list of known solutions in general relativity and its extensions. This family of BHs reduces to known solutions if some of its parameters are adjusted to specific values.

The thermodynamics of these BHs follows closely that of two-parameter Taub-NUT family, whith mass, temperature, and thermodynamic potential (associated with the gravitational Misner charge) acquiring a Lorentz-symmetry-breaking-parameter ($\ell$) dependence that is inversely proportional to $\sqrt{1+\ell}$.

The thermodynamic topological classification of these three-parameter Taub-NUT-like BHs is that of the two-parameter Taub-NUT BHs.

%-------------------------------------------------------------
%\begin{acknowledgments}

%\end{acknowledgments}

%\bibliography{reference.bib}

\begin{thebibliography}{999}

\bibitem{Colladay:1998fq}
D.~Colladay and V.~A.~Kostelecky,
\href{https://journals.aps.org/prd/abstract/10.1103/PhysRevD.58.116002}
{Phys. Rev. D \textbf{58} (1998), 116002}

\bibitem{Kostelecky:1991ak}
V.~A.~Kostelecky and R.~Potting,
\href{https://www.sciencedirect.com/science/article/abs/pii/0550321391900715?via%3Dihub}
{Nucl. Phys. B \textbf{359} (1991), 545-570}

\bibitem{Kostelecky:1989jp}
V.~A.~Kostelecky and S.~Samuel,
\href{https://journals.aps.org/prl/abstract/10.1103/PhysRevLett.63.224}
{Phys. Rev. Lett. \textbf{63} (1989), 224}

\bibitem{Gambini:1998it}
R.~Gambini and J.~Pullin,
\href{https://journals.aps.org/prd/abstract/10.1103/PhysRevD.59.124021}
{Phys. Rev. D \textbf{59} (1999), 124021}

\bibitem{Carroll:2001ws}
S.~M.~Carroll, J.~A.~Harvey, V.~A.~Kostelecky, C.~D.~Lane and T.~Okamoto,
\href{https://journals.aps.org/prl/abstract/10.1103/PhysRevLett.87.141601}
{Phys. Rev. Lett. \textbf{87} (2001), 141601}


\bibitem{Kostelecky:1989jw}
V.~A.~Kostelecky and S.~Samuel,
\href{https://journals.aps.org/prd/abstract/10.1103/PhysRevD.40.1886}
{Phys. Rev. D \textbf{40} (1989), 1886-1903}


\bibitem{Casana:2017jkc}R. Casana and A. Cavalcante, \href{https://doi.org/10.1103/PhysRevD.97.104001}{Phys. Rev. D \textbf{97}, 104001 (2018)}


\bibitem{Maluf:2020kgf}
R.~V.~Maluf and J.~C.~S.~Neves,
\href{https://journals.aps.org/prd/abstract/10.1103/PhysRevD.103.044002}
{Phys. Rev. D \textbf{103} 
	(2021) no.4, 044002}

\bibitem{Ovgun:2018xys}
A.~{\"O}vg{\"u}n, K.~Jusufi and {\.I}.~Sakall{\i},
\href{https://journals.aps.org/prd/abstract/10.1103/PhysRevD.99.024042}
{Phys. Rev. D \textbf{99} (2019) no.2, 024042}

\bibitem{Gullu:2020qzu}
{\.I}.~G{\"u}ll{\"u} and A.~{\"O}vg{\"u}n,
\href{https://linkinghub.elsevier.com/retrieve/pii/S0003491621003237}
{Annals Phys. \textbf{436} (2022), 168721}

\bibitem{Ding:2021iwv}
C.~Ding, X.~Chen and X.~Fu,
\href{https://linkinghub.elsevier.com/retrieve/pii/S0550321322000396}
{Nucl. Phys. B \textbf{975} (2022), 115688}

\bibitem{Adailton23}
A.A. Ara\'{u}jo Filho, J.R. Nascimento, A.Yu. Petrov and P.J. Porf\'{\i}rio,
\href{https://doi.org/10.1103/PhysRevD.108.085010}
{Phys. Rev. D \textbf{108}, 085010 (2023)}

\bibitem{Poulis:2021nqh}
F.~P.~Poulis and M.~A.~C.~Soares,
\href{https://link.springer.com/article/10.1140/epjc/s10052-022-10547-y}
{Eur. Phys. J. C \textbf{82} (2022) no.7, 613}

\bibitem{Adailton24}
A.A. Ara\'{u}jo Filho, J.R. Nascimento, A.Yu. Petrov and P.J. Porf\'{\i}rio,
\href{https://doi.org/10.1088/1475-7516/2024/07/004}
{JCAP07(2024)004}

\bibitem{Liu:2022dcn}
W.~Liu, X.~Fang, J.~Jing and J.~Wang,
\href{https://link.springer.com/article/10.1140/epjc/s10052-023-11231-5}
{Eur. Phys. J. C \textbf{83} (2023) no.1, 83}

\bibitem{Mai:2023ggs}
Z.~F.~Mai, R.~Xu, D.~Liang and L.~Shao,
\href{https://journals.aps.org/prd/abstract/10.1103/PhysRevD.108.024004}
{Phys. Rev. D \textbf{108} (2023) no.2, 024004}

\bibitem{Yang:2023wtu}
K.~Yang, Y.~Z.~Chen, Z.~Q.~Duan and J.~Y.~Zhao,
\href{https://journals.aps.org/prd/abstract/10.1103/PhysRevD.108.124004}
{Phys. Rev. D \textbf{108} (2023) no.12, 124004}

\bibitem{Xu:2023xqh}
R.~Xu, D.~Liang and L.~Shao,
\href{https://iopscience.iop.org/article/10.3847/1538-4357/acbdfb}
{Astrophys. J. \textbf{945} (2023) no.2, 148}

\bibitem{Zhang:2023wwk}
X.~Zhang, M.~Wang and J.~Jing,
\href{https://link.springer.com/article/10.1007/s11433-023-2153-6}
{Sci. China Phys. Mech. Astron. \textbf{66} (2023) no.10, 100411}

\bibitem{Wang:2021gtd}
Z.~Wang, S.~Chen and J.~Jing,
\href{https://link.springer.com/article/10.1140/epjc/s10052-022-10475-x}
{Eur. Phys. J. C \textbf{82} (2022) no.6, 528}

\bibitem{Taub:1950ez}
A.~H.~Taub,
\href{https://www.jstor.org/stable/1969567?origin=crossref}
{Annals Math. \textbf{53} (1951), 472-490}

\bibitem{Newman:1963yy}
E.~Newman, L.~Tamburino and T.~Unti,
\href{https://pubs.aip.org/aip/jmp/article-abstract/4/7/915/230250/Empty-Space-Generalization-of-the-Schwarzschild?redirectedFrom=fulltext}
{J. Math. Phys. \textbf{4} (1963), 915}

\bibitem{Misner:1963fr}
C.~W.~Misner,
\href{https://pubs.aip.org/aip/jmp/article-abstract/4/7/924/230241/The-Flatter-Regions-of-Newman-Unti-and-Tamburino-s?redirectedFrom=fulltext}
{J. Math. Phys. \textbf{4} (1963), 924-938}

\bibitem{eq1}M. Azreg-A\"{\i}nou, \href{https://doi.org/10.1140/epjc/s10052-015-3258-3}{Eur. Phys. J. C \textbf{75}, 34 (2015)}

\bibitem{eq2}M. Azreg-A\"{\i}nou, \href{https://doi.org/10.1103/PhysRevD.91.064049}{Phys. Rev. D \textbf{91}, 064049 (2015)}

\bibitem{eq3}R.A. Hennigar, D. Kubiz\v{n}\'{a}k and R.B. Mann, \href{https://doi.org/10.1103/PhysRevD.100.064055}{Phys. Rev. D \textbf{100}, 064055 (2019)}

\bibitem{Wu:2019pzr}
S.~Q.~Wu and D.~Wu,
\href{https://journals.aps.org/prd/abstract/10.1103/PhysRevD.100.101501}
{Phys. Rev. D \textbf{100}, 101501 (2019)}

\bibitem{Liu:2022wku}
H.~S.~Liu, H.~Lu and L.~Ma,
\href{https://link.springer.com/article/10.1007/JHEP10(2022)174}
{JHEP \textbf{10}, 174 (2022)}

\bibitem{eq4}J.F. Liu and H.S. Liu, \href{https://doi.org/10.1140/epjc/s10052-024-12826-2}{Eur. Phys. J. C \textbf{84}, 515 (2024)}

%\bibitem{Hennigar:2019ive}
%R.~A.~Hennigar, D.~Kubiz{\v{n}}{\'a}k and R.~B.~Mann,
%\href{https://journals.aps.org/prd/abstract/10.1103/PhysRevD.100.064055}
%{Phys. Rev. D \textbf{100} (2019) no.6, 064055}

%\bibitem{Liu:2023uqf}
%J.~F.~Liu and H.~S.~Liu,
%\href{https://link.springer.com/article/10.1140/epjc/s10052-024-12826-2}
%{Eur. Phys. J. C \textbf{84} (2024) no.5, 515}

\bibitem{Chen:2024knw}
Y.~Q.~Chen, H.~S.~Liu and H.~Lu,
\href{https://journals.aps.org/prd/abstract/10.1103/PhysRevD.110.104068}
{Phys. Rev. D \textbf{110}, 104068 (2024)}

\bibitem{LIGOScientific:2017vwq}
B.~P.~Abbott \textit{et al.} [LIGO Scientific and Virgo],
\href{https://journals.aps.org/prl/abstract/10.1103/PhysRevLett.119.161101?__cf_chl_rt_tk=zySmidEi4Qaoi.ih.C0_sl84rC9Nk.PeVqa2LI9yhas-1758385784-1.0.1.1-bXysEKWqfTqYr9sYeWmSsV8rH79HNctV2LI0F8hlYXA}
{Phys. Rev. Lett. \textbf{119} (2017) no.16, 161101}

\bibitem{LIGOScientific:2017zic}
B.~P.~Abbott \textit{et al.} [LIGO Scientific, Virgo, Fermi-GBM and INTEGRAL],
\href{https://iopscience.iop.org/article/10.3847/2041-8213/aa920c}
{Astrophys. J. Lett. \textbf{848} (2017) no.2, L13}

\bibitem{EventHorizonTelescope:2019dse}
K.~Akiyama \textit{et al.} [Event Horizon Telescope],
\href{https://iopscience.iop.org/article/10.3847/2041-8213/ab0ec7}
{Astrophys. J. Lett. \textbf{875} (2019), L1}

\bibitem{EventHorizonTelescope:2019uob}
K.~Akiyama \textit{et al.} [Event Horizon Telescope],
\href{https://iopscience.iop.org/article/10.3847/2041-8213/ab0c96}
{Astrophys. J. Lett. \textbf{875} (2019) no.1, L2}

\bibitem{EventHorizonTelescope:2019jan}
K.~Akiyama \textit{et al.} [Event Horizon Telescope],
\href{https://iopscience.iop.org/article/10.3847/2041-8213/ab0c57}
{Astrophys. J. Lett. \textbf{875} (2019) no.1, L3}

\bibitem{EventHorizonTelescope:2019ths}
K.~Akiyama \textit{et al.} [Event Horizon Telescope],
\href{https://iopscience.iop.org/article/10.3847/2041-8213/ab0e85}
{Astrophys. J. Lett. \textbf{875} (2019) no.1, L4}

\bibitem{EventHorizonTelescope:2019pgp}
K.~Akiyama \textit{et al.} [Event Horizon Telescope],
\href{https://iopscience.iop.org/article/10.3847/2041-8213/ab0f43}
{Astrophys. J. Lett. \textbf{875} (2019) no.1, L5}

\bibitem{EventHorizonTelescope:2019ggy}
K.~Akiyama \textit{et al.} [Event Horizon Telescope],
\href{https://iopscience.iop.org/article/10.3847/2041-8213/ab1141}
{Astrophys. J. Lett. \textbf{875} (2019) no.1, L6}

\bibitem{Chakraborty:2019rna}
C.~Chakraborty and S.~Bhattacharyya,
\href{https://iopscience.iop.org/article/10.1088/1475-7516/2019/05/034}
{JCAP \textbf{05} (2019), 034}

\bibitem{Ghasemi-Nodehi:2021ipd}
M.~Ghasemi-Nodehi, C.~Chakraborty, Q.~Yu and Y.~Lu,
\href{https://link.springer.com/article/10.1140/epjc/s10052-021-09696-3}
{Eur. Phys. J. C \textbf{81} (2021) no.10, 939}

\bibitem{Jafarzade:2025zbg}
k.~Jafarzade, M.~Ghasemi-Nodehi, F.~Sadeghi and B.~Mirza,
\href{https://iopscience.iop.org/article/10.1088/1475-7516/2025/03/063}
{JCAP \textbf{03} (2025), 063}

\bibitem{Chakraborty:2017nfu}
C.~Chakraborty and S.~Bhattacharyya,
\href{https://journals.aps.org/prd/abstract/10.1103/PhysRevD.98.043021}
{Phys. Rev. D \textbf{98} (2018) no.4, 043021}

\bibitem{Chakraborty:2022ltc}
C.~Chakraborty and S.~Bhattacharyya,
\href{https://journals.aps.org/prd/abstract/10.1103/PhysRevD.106.103028}
{Phys. Rev. D \textbf{106} (2022) no.10, 103028}

\bibitem{Chakraborty:2023wpz}
C.~Chakraborty and B.~Mukhopadhyay,
\href{https://link.springer.com/article/10.1140/epjc/s10052-023-12070-0}
{Eur. Phys. J. C \textbf{83} (2023) no.10, 937}

\bibitem{Chen25}
Y.-Q. Chen and H.-S. Liu,
\href{https://doi.org/10.1103/wmhj-83x3}
{Phys. Rev. D \textbf{112} (2025) no.08, 084040}
%\href{https://arxiv.org/pdf/2505.23104}
%{arXiv:2505.23104v1 [gr-qc]}

%\bibitem{Ding:2019mal}C. Ding, C. Liu, R. Casana and A. Cavalcante, \href{https://doi.org/10.1140/epjc/s10052-020-7743-y}{Eur. Phys. J. C \textbf{80}, 178 (2020)}

\bibitem{mass1}A. Ashtekar and S. Das, \href{https://doi.org/10.1088/0264-9381/17/2/101}{Class. Quantum Grav. \textbf{17}, L17 (2000)}

\bibitem{mass2}N. Okuyama and J. Koga, \href{https://doi.org/10.1103/PhysRevD.71.084009}{Phys. Rev. D \textbf{71}, 084009 (2005)}

\bibitem{mass3}J.J. Peng, C.L. Zou and H.F. Liu, \href{https://doi.org/10.1088/1402-4896/ac1cd1}{Phys. Scr. \textbf{96}, 125207 (2021)}	

\bibitem{Poincare}M. Azreg-A\"{\i}nou and M.E. Rodrigues, \href{https://doi.org/10.1007/JHEP09(2013)146}{JHEP09, 146 (2013)}

\bibitem{Jiayue23}J. Yang, R.B. Mann, 
\href{https://doi.org/10.1007/JHEP08%282023%29028}{JHEP08(2023)028}

\bibitem{ettbh}S.P. Wu, S.-J. Yang, S.-W. Wei, 
\href{https://arxiv.org/abs/2508.01614v1}{	arXiv:2508.01614 [hep-th]}

\bibitem{AR25}M. Azreg-A\"{\i}nou and M. Rizwan, \href{https://doi.org/10.1088/1475-7516/2025/05/091}{JCAP05, 091 (2025)}

\bibitem{Poincare1}H. Poincar\'{e}, \href{https://www.numdam.org/item/?id=JMPA_1881_3_7__375_0}{Journal de math\'{e}matiques pures et appliqu\'{e}es \textbf{7}(3), 375-422 (1881)}

\bibitem{Poincare2}H. Poincar\'{e}, \href{https://www.numdam.org/item/JMPA_1885_4_1__167_0/}{Journal de math\'{e}matiques pures et appliqu\'{e}es \textbf{1}(4), 167-244 (1885)}

\bibitem{Hopf}P. Alexandroff and H Hopf, \textit{Topologie. I.}, Berichtigter Reprint, Die Grundlehren der mathematischen
Wissenschaften, Band 45, Springer--Verlag, Berlin--New York (1974)

\bibitem{book}J. Milnor, \textit{Morse Theory}, Annals of Mathematics Studies Book 51, Princeton University Press, Princeton, NJ (1963)

\bibitem{class}S.-W. Wei, Y.-X. Liu and R.B. Mann, \href{https://doi.org/10.1103/PhysRevD.110.L081501}{Phys. Rev. D \textbf{110}, L081501 (2024)}

\bibitem{York}J.W. York Jr., \href{https://doi.org/10.1103/PhysRevD.33.2092}{Phys. Rev. D \textbf{33}, 2092 (1986)}



\end{thebibliography}
%\bibliographystyle{apsrev4-1}

\end{document}